# Co-resonant enhancement of spin-torque critical currents in spin-valves with synthetic-ferrimagnet free-layer

Neil Smith, Stefan Maat, Matthew J. Carey, Jeffrey R. Childress.

San Jose Research Center, Hitachi Global Storage Technologies, San Jose, CA 95135

It is experimentally shown that the critical current for onset of spin-torque instability in current-perpendicular-to-plane spin-valves can be strongly enhanced using "synthetic ferrimagnet" free-layers of form $FM_1/Ru/FM_2$ (FM=ferrromagnet). However, this enhancement occurs for only one polarity of bias current. A two-macrospin model is shown to reproduce the observations. The model suggests that this phenomenon is related to a polarity-dependent, spin-torque induced co-resonance between the two natural dynamic modes of the $FM_1/FM_2$ couple. The resonance condition facilitates energy transfer out of the spin-torque destabilized mode into the other stable mode whose effective damping is actually enhanced by spin-torques, thereby delaying the onset of instability of this coupled system to larger critical currents.

Spin-torque phenomena, as manifested in giant-magnetoresistive spin-valves film stacks lithographically patterned into ~100 nm nanopillars and driven with electrical currents perpendicular to the film plane have in recent years been the active study of numerous theoretical and experimental papers,[1] both for their novel physics as well as potential applications for magnetic memory elements, microwave oscillators, and magnetic recording read heads. In essentially all of these studies, the dynamically active magnetic layer, or "free layer" of the spin-valve film stack, is either theoretically modeled or experimentally fabricated as a single ferromagnetic layer. This paper investigates, through both experimental measurement and theoretical modeling, the novel spin-torque dynamics of a "synthetic-ferrimagnetic" free-layer of the form FM1/Ru/FM2, consisting of two ferromagnetic (FM) films of unequal thickness $t_{FM1} \gg t_{FM2}$ separated by a thin (0.8 nm) Ru spacer which promotes well-known[2], strong antiparallel coupling between the two FM layers. Compared to the simple free-layer system, the FM1/Ru/FM2 couple has two additional spin-torque-producing (Ru/FM) interfaces, and permits (in the simple macrospin picture) two independent, non-degenerate, natural modes of oscillation. As will be discussed below, these features can lead to a novel condition of spin-torque-induced "quasi-co-resonance" of these two modes which greatly impacts the spin-torque-stability of such devices, and which carries potentially important practical implications for their use in the aforementioned applications.

The present experiments use multilayer films of form AFM/PL/Ru/RL/Cu/FL1/Ru/FL2 (excluding seed and cap layers). The first ferromagnetic pinned-layer (PL) is exchange-pinned to the antiferromagnetic (AFM) layer, and is also strongly antiparallel-coupled to a second FM reference-layer (RL) across a thin Ru spacer. The PL and RL layers are closely moment-matched, forming a "synthetic-antiferromagnetic" couple (as is common practice for such devices) which consequently does not respond to a modest external magnetic fields. More unique to the present structures, the first free-layer (FL1) is also antiparallel-coupled to a second free-layer (FL2), forming the "synthetic-ferrimagnetic-free-layer" (SFM-FL) with sheet-film $M$-$H$ behavior (at modest external fields) equivalent to a single FM film of thickness $t_{FL1} - t_{FL2}$.

A first set of experimental measurements, described in Fig. 1, uses NiFe free-layers with $t_{FL1} = 4\,\text{nm} + t_{FL2}$ including a control with $t_{FL2} = 0$, and two SFM-FL designs with $t_{FL2} = 2\,\text{nm}$ and $3\,\text{nm}$. The devices tested have been patterned into 75-nm circular pillars using E-beam lithography.[3] Resistance ($R$-$H$) loops are measured (at -5mV bias) in fields collinear ($H_x$) and transverse ($H_y$) to the IrMn pinning direction. All chosen devices have nonhysteretic, square $R$-$H_x$ (for $|H_x| \leq 1\,\text{kOe}$) and near-symmetric (about $H_y = 0$) $R$-$H_y$. Accompanying each $R$-$H$ data set are two $N$-$I_e$ loops, which measure narrow-band noise *N* vs *electron* current $I_e$ with constant applied fields of either $H_x \cong +600\,\text{Oe}$ or -600 Oe to align FL1 magnetization either antiparallel (AP) or parallel (P) with that of the RL. Positive electron current travels from RL to FL (Fig. 1a). The current is driven by a 2-Hz sawtooth generator with sync pulse triggering the 0.5-sec sweep of a (zero-span) spectrum analyzer. The $N$-$I_e$ loops are averaged over $\approx 50$ sweeps. The $I_e \cong 0$ electronics noise ($\sim 0.8\,\text{nV}/\sqrt{\text{Hz}}$) is subtracted out.

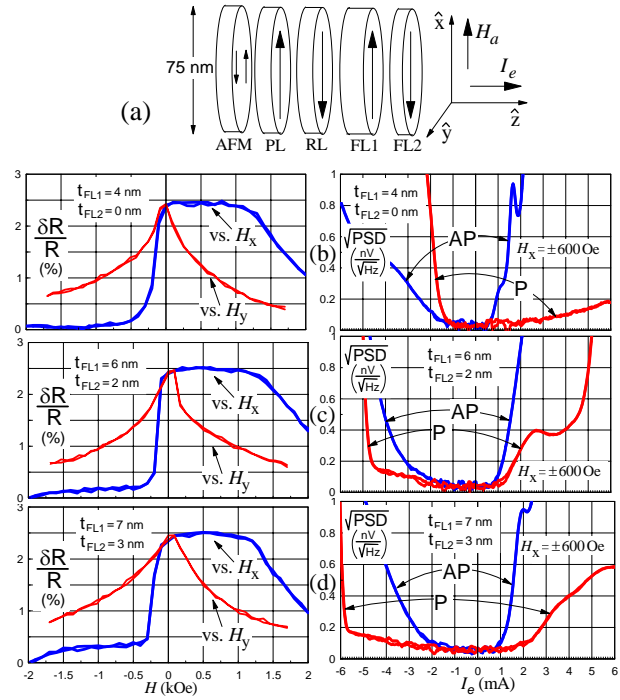

Fig. 1. (a) cartoon of device geometry. (b-d) *R-H* loops (as % δR/R) and *N-$I_e$* loops (as rms power spectral density at 75 MHz); for $t_{FL2}$ = 0 (b), 2nm (c), and 3nm (d). Spin-valve stack structure: IrMn(7)/CoFe(3)a/Ru(0.7)/CoFe(3)/Cu(4)/ NiFe(4+ $t_{FL2}$)/Ru(0.7)/NiFe($t_{FL2}$); ( ) denotes film thickness in nm.

The $N$-$I_e$ technique[4] measures the 1/*f*-like noise associated with thermal perturbations of well-known precessional motion of a unidirectionally stable FL once spin-torque instability begins.[5]. This onset is readily observed by the sharp increase in noise above the $\approx 0.03\,\text{nV}/\sqrt{\text{Hz}}$ /mA residual electronics noise for these $\approx 11\Omega$ devices. The "critical currents", $I_e^{\text{crit}}$, for this onset are found by simple inspection. The $I_e^{\text{crit}}$ were typically insensitive to few hundred Oe variations in $H_x$.

For all thicknesses of FL2, there is an observed AP-state *negative* critical point $-I_{e\,\text{AP}}^{\text{crit}} \approx 2\text{-}2.5\,\text{mA}$ which was previously shown to be spin-torque-instability of the RL/PL[5,6]. The SFM-FL devices alone show an additional *positive* critical point $+I_{e\,\text{P}}^{\text{crit}}$ in the P-state, which is discussed further below. For the

$t_{FL2} = 0$ control, the polarity asymmetry ratio $(-I_{eP}^{crit})_{FL1} /(+I_{eAP}^{crit})_{FL1} \approx 2.5\text{-}3$ is similar to earlier observations,[4-6] and is a known consequence of the intrinsic angular dependence of electrical transport in these all-metallic devices. It is unrelated to the unexpected polarity asymmetry discussed below.

Fig. 2 shows a summary of similar $N\text{-}I_e$ measurements for a second, modified film stack which includes thin CoFe layers at the Cu/FL1 and FL1,2/Ru interfaces.[7] Referenced to $Ni_{80}Fe_{20}$ films of equal moment, the magnetic thicknesses of FL1,2 are similar to those shown in Fig. 1.

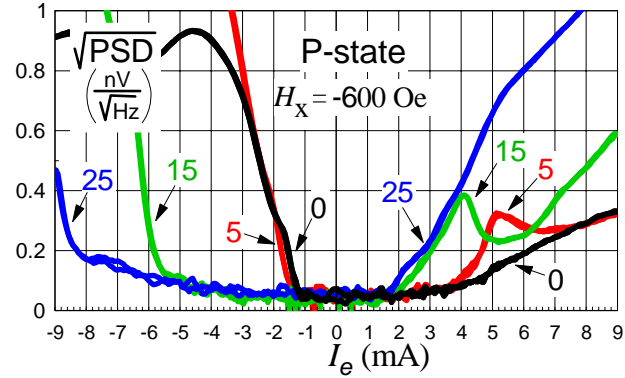

Fig. 2. P-state $N\text{-}I_e$ loops (rms power spectral density); for spin-valve stack::
IrMn(7)/CoFe(3)/Ru(0.6)/CoFe(3)/Cu(5)/CoFe(0.6)NiFe(4+ $t_{FL2}$) /CoFe(0.2)/Ru(0.6)/CoFe(0.2)/NiFe($t_{FL2}$);
( ) denotes film thickness in nm.. In $Ni_{80}Fe_{20}$ equivalent Angstroms thickness, $t_{FL1} = t_{FL2}+45$, $t_{FL2}$ as indicated.

Fig. 3 summarizes the experimental results for $I_e^{crit}$ for both stack structures, which includes 3 or 4 devices for each value of $t_{FL2}$, for which the data in Figs. 1 and 2 are representative.

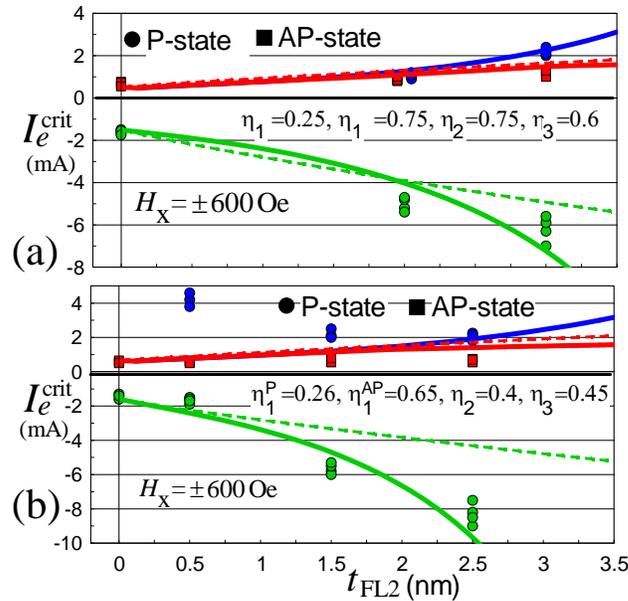

Fig. 3. Critical current vs. $t_{FL2}$ for film stacks from Fig. 1 (a), and Fig. 2 (b). Solid circles (squares) for P-state (AP-state). Solid curves are model results as described in text, using η-coefficients as indicated in figures. Dashed curves are for $η_2 = η_3 = 0$, but which exclude $I_{eP}^{crit} > 0$.

In both cases, *there is a striking increase in the magnitude of* $-I_{eP}^{crit}$ *with increasing* $t_{FL2}$, *which is even more dramatically demonstrated for the second, CoFe-interfaced FL stack structure*. In stark contrast to these observations regarding $-I_{eP}^{crit}$, it is seen that $+I_{eAP}^{crit}$ shows rather little absolute change with $t_{FL2}$. These two counter-intuitive results agree well with analogous $dV/dI$ measurements[7] using test devices (same film stack as in Fig. 2) of similar area but asymmetric non-circular geometry.

For a theoretical insight into this phenomenon, we start with a simple two-macrospin model for FL1/FL2 which treats the RL/PL as an inert, spin-current polarizer. Magnetostatic, Ru-coupling and Zeeman terms for the system free energy $E$ are taken to be

$$E/(M_sV)_{FL1} = \frac{1}{2}\sum_{u;\,j,k=1}^{2} m_{ju}H_{jk}^u m_{ku} + H_{ru}\hat{m}_1\cdot\hat{m}_2 \quad (1)$$
$$-H_a[m_{1x} + (M_st)_{FL2}/(M_st)_{FL1}\, m_{2x}]$$

where $\hat{m}_{1,2}$ are the unit magnetization vectors for FL1,2, the set of tri-indicied $H_{jk=1,2}^{u=x,y,z}$ are magnetostatic energy coefficients, $J_{ru}$ is the interfacial coupling strength, $H_{ru}\equiv J_{ru}/(M_st)_{FL1}$, $\boldsymbol{H}_a = H_a\hat{x}$ is the applied field, and $V_{FL1}$ is the volume of FL1. Slonczewski-type[8] spin-torques $\boldsymbol{\tau}=(M_sV)_{FL1}\boldsymbol{H}^{ST}\times\hat{m}$ at the Cu/FL1, FL1/Ru, and Ru/FL2 interfaces are included as follows:

$$\boldsymbol{H}_1^{ST} = \eta_1 H_{ST}\hat{m}_1\times(\pm\hat{x}) + \eta_2 H_{ST}\hat{m}_2\times\hat{m}_1$$
$$\boldsymbol{H}_2^{ST} = \eta_3 H_{ST}\hat{m}_2\times\hat{m}_1;\ H_{ST}\equiv(\hbar/2e)I_e/(M_sV)_{FL1} \quad (2)$$
$$\boldsymbol{H}_{j=1,2}^{eff} = -1/(M_sV)_{FL1}\partial E/\partial\hat{m}_j + \boldsymbol{H}_j^{ST}$$

to form the total effective field $\boldsymbol{H}_{j=1,2}^{eff}$. In (2), $\pm\hat{x}$ refers to $\hat{m}_{RL}$ (P and AP-states), and the $I_e=0$ equilibrium state now defined to be $\hat{m}_{01}=\hat{x}$, $\hat{m}_{02}=-\hat{x}$. The $\eta$-coefficients will be discussed below.

The additional (two) degrees of freedom of the second macrospin substantially complicates the algebraic description relative to the well-known 1-macrospin case.. As described previously,[5,9] individual, local $x'y'z'$ coordinates where $\hat{m}_{01}=\hat{x}'=\hat{m}_{02}$ are used to construct the following matrix formulation of the *linearized* Gilbert equations of motion for the *two*-dimensional vectors $\boldsymbol{m}'_{j=1,2}=(m_{jy'},m_{jz'})$:

$$(\ddot{\boldsymbol{D}}+\ddot{\boldsymbol{G}}) \cdot \frac{d\boldsymbol{m}'}{dt} + \ddot{\boldsymbol{H}} \cdot \boldsymbol{m}'(t) = 0, \quad \boldsymbol{m}' \equiv \begin{pmatrix} \boldsymbol{m}'_1 \\ \boldsymbol{m}'_2 \end{pmatrix}$$

$$\ddot{G}_{jk} \equiv \frac{t_j}{\gamma t_1} \begin{pmatrix} 0 & -1 \\ 1 & 0 \end{pmatrix} \delta_{jk}, \quad \ddot{D}_{jk} \equiv \frac{\alpha t_j}{\gamma t_1} \begin{pmatrix} 1 & 0 \\ 0 & 1 \end{pmatrix} \delta_{jk}$$

$$H_{jk}^{u'v'} \equiv (\boldsymbol{H}_j^{\text{eff}} \cdot \hat{\boldsymbol{m}}_j) \delta_{jk} - \sum_{u,v} \frac{\partial m'_{ju'}}{\partial m_{ju}} \frac{\partial H_{ju}^{\text{eff}}}{\partial m_{kv}} \frac{\partial m_{kv}}{\partial m'_{kv'}}$$

$$\ddot{H}_{11} \equiv \begin{pmatrix} H_5 & -\eta' H_{\text{ST}} \\ \eta' H_{\text{ST}} & H_6 \end{pmatrix}, \quad \ddot{H}_{12} \equiv \begin{pmatrix} -H_3 & -\eta_2 H_{\text{ST}} \\ -\eta_2 H_{\text{ST}} & H_4 \end{pmatrix}$$

$$\ddot{H}_{21} \equiv \begin{pmatrix} -H_3 & \eta_3 H_{\text{ST}} \\ \eta_3 H_{\text{ST}} & H_4 \end{pmatrix}, \quad \ddot{H}_{22} \equiv \begin{pmatrix} H_7 & \eta_3 H_{\text{ST}} \\ -\eta_3 H_{\text{ST}} & H_8 \end{pmatrix}$$

$$\eta' \equiv \pm \eta_1 + \eta_2, \quad H_3 \equiv H_{\text{ru}} + H_{12}^y, \quad H_4 = H_{\text{ru}} + H_{12}^z$$

$$H_5 \equiv H_3 + H_a, \quad H_6 \equiv H_5 + H_{11}^z - H_{11}^y$$

$$H_7 \equiv H_3 - (t_1/t_2) H_a, \quad H_8 \equiv H_7 + H_{22}^z - H_{22}^y$$

(3)

where gyromagnetic ratio $\gamma \cong 19\,\text{Mrad/(sec-Oe)}$, and $\alpha > 0$ is the Gilbert damping parameter. In (2), $\ddot{\boldsymbol{H}} \Leftrightarrow H_{jk}^{u'v'}$ is a $4\times 4$ tensor-matrix formed from the 2D Cartesian tensors $\ddot{H}_{jk}$ given explicit in (3), and similar for $\ddot{\boldsymbol{D}}$ and $\ddot{\boldsymbol{G}}$. The expressions for $H_{5,7}$ assume the symmetry $H_{jk}^x = H_{jk}^y$.

The natural modes for this system are nontrivial solutions of (3) of the form $\boldsymbol{m}'(t) \propto e^{-st}$. The 4 roots $s = \sigma_1 \pm i\omega_1, \sigma_2 \pm i\omega_2$ satisfy $\det|\ddot{\boldsymbol{H}} - s(\ddot{\boldsymbol{D}}+\ddot{\boldsymbol{G}})| = 0$, but are more generally found from the eigenvalues of the matrix $\ddot{\boldsymbol{H}} \cdot (\ddot{\boldsymbol{D}}+\ddot{\boldsymbol{G}})^{-1}$.[9] Only two of these roots, $s_{1,2} = \sigma_{1,2} + i\omega_{1,2}$, describe physically distinct modes. The spin-torque terms in (2) yield a nonreciprocal $\ddot{\boldsymbol{H}}$ (i.e., $H_{jk}^{u'v'} \neq H_{kj}^{v'u'}$), which can be shown[10] to permit *unstable* modes with $\text{Re}(s) < 0$. This corresponds physically to the spin-torque-instability seen in the $N$-$I_e$ of Figs. 1,2. Computing $s_{1,2}(I_e; t_{\text{FL2}})$, the *least* value of $\pm I_e$ where $\sigma_< \equiv \min(\sigma_1, \sigma_2) \to 0$ yields the model-predicted $I_e^{\text{crit}}(t_{\text{FL2}})$ curves shown in Fig. 3.

The computed $I_e^{\text{crit}}$ vs. $t_{\text{FL2}}$ assumed a NiFe saturation magnetization $M_s = 800\,\text{emu/cc}$, AF/Ru-coupling $J_{\text{ru}} = 1\,\text{erg/cm}^2$, $H_a = 600\,\text{Oe}$, and $\alpha = 0.02$. The $H_{jk}^u$ were computed (analytically) from the interaction energy between uniformly magnetized rectangular solids ($\cong 65$ nm squares with area equal to 75 nm circles). Values of $\eta_1^P$, $\eta_1^{AP}$ describe the spin-torque amplitude at the Cu/FL1 interface (in P or AP states), and were chosen to match the measured $I_e^{crit}$ for control devices with $t_{FL2} \to 0$. The P-state values are close to the theoretical estimate[5,8] $\eta_1^P \approx \beta/2$, with $\beta \approx 0.65$ roughly the mean spin-polarization between NiFe and CoFe.[11]

The (non-unique) values of $\eta_2$ and $\eta_3$, which govern the spin-torque amplitude at the FL1/Ru and Ru/FL2 interfaces, were chosen to obtain an eyeball fit with all other $I_e^{crit}$ vs. $t_{FL1,2}$ data. The good agreement between data and model in Fig. 3 is obtained with constant $\eta_{1-3}$ coefficients that are independent of $t_{FL1,2}$ ($\eta_{2,3}$ also independent of P or AP), while using a simple magnetic model with some imprecisely known parameter values, e.g., $J_{ru}$ and $\alpha$. The level of agreement is perhaps fortuitous, though the essential physics is believed robust to any shortcomings of the model.

An $\eta_2 = \eta_3 = 0$ model can perhaps provide a plausible fit to the $I_{eP}^{crit} < 0$ and $I_{eAP}^{crit} > 0$ NiFe-FL data in Fig. 3a, but this is *not* so for the CoFe/NiFe/CoFe-FL data of Fig. 3b. In either case, it is incapable of accounting for the SFM-FL data with $I_{eP}^{crit} > 0$, which originates from spin-torque-instability at the Ru/FL2 interface. Fitting the full model to this data required a value $\eta_3 \geq 0.4$. In addition, the model fails for the $I_{eP}^{crit} > 0$ data in Fig. 3b at $t_{FL2} = 0.5$ nm, which may reflect excess damping as $t_{FL2} \to 0$, and/or a breakdown (due to finite *transverse* spin-diffusion length[12]) of the purely interfacial form of spin-torque implicitly assumed in (2).

In the limit $J_{ru} \to \infty \Rightarrow m_{2y'} \to m_{1y'}, m_{2z'} \to -m_{1z'}$ one may obtain from (3) the following analytical solutions for $I_e^{crit}$:

$$I_{e(P\,or\,AP)}^{crit} \to \left[ \frac{t_{FL1} + t_{FL2}}{t_{FL1} - t_{FL2}} \right] \frac{\alpha (M_s V)_{FL1}}{\hbar/2e} \frac{H_0}{(-\eta_1^P \,or\, +\eta_1^{AP})}$$

$$H_0 \equiv (1 - t_{FL2}/t_{FL1})H_a + \frac{1}{2} \sum_{j,k=1}^{2} (-1)^{j+k} (H_{jk}^z - H_{jk}^y)$$

(4)

which also applies when $t_{FL2} = 0$. With $t_{FL1} - t_{FL2}$ fixed, (4) indicates (excluding any $t_{FL}$-dependence of $H_0$) a $(t_{FL1} + t_{FL2}) \cdot t_{FL1}$ scaling, *but equally so for either* $-I_{eP}^{crit}$ *or* $+I_{eAP}^{crit}$. The latter sharply contradicts experiment. Further, (4) excludes the additional observed cases of $I_{eP}^{crit} > 0$ when

$t_{FL2} > 0$. The underlying physics behind the *asymmetric*, strong superlinear (or weak) $t_{FL2}$-dependence of $I_{eP}^{crit} < 0$ (or $I_{eAP}^{crit} > 0$) would thus appear connected with the finiteness of $J_{ru}$.

This is further elucidated in Fig.4. Computed as a continuous function of $J_{ru}$ are critical currents $I_e^{crit}(J_{ru})$, as well as natural-mode parameters $\Delta f \equiv \frac{1}{2\pi}(\omega_> - \omega_<)$ and $\sigma_> \equiv \max(\sigma_1, \sigma_2)$ evaluated *at* $I_e = I_e^{crit}(J_{ru})$. In the case of *negative* $I_{eP}^{crit} < 0$ in Figs. 4a,b, the spin-torque terms radically alter the natural oscillation frequencies $\omega_1, \omega_2$, even to the point of inducing a literal *co-resonance*, i.e., $\Delta f \equiv \frac{1}{2\pi}(\omega_> - \omega_<) \to 0$, between the two modes at a finite $J_{ru}$. Closely accompanying the co-resonance is a broad peak in both $-I_{eP}^{crit}(J_{ru})$ and $\sigma_>(J_{ru})$, with a *large* maximum at $J_{ru} \equiv J_{ru}^{max}$.

The $\sigma_>$ are the temporal decay rates (or line-widths) of the *stable* mode. Here, spin-torques can *increase* the rate of energy loss from that mode of oscillation well beyond that of intrinsic damping (e.g., $\sigma_> \approx 4 \, \text{Gsec}^{-1}$ at $I_e = 0$.). The third (damping) matrix $\vec{\vec{D}}$ in (3), along with nonreciprocal spin-torque contributions to $\vec{\vec{H}}$, imply that the two natural modes are *non*-orthogonal, and are coupled both by spin-torques and weakly by intrinsic damping. This dynamic coupling allows energy transfer between modes, which is further strongly enhanced at and near the condition of co-resonance. However, this enhancement does not strictly require $\Delta f \to 0$, but can occur under more general conditions of "quasi-co-resonance" where $\Delta f / \sigma_> < 1$, i.e., when the difference in the modes' resonant frequencies is smaller than the effective line-width $\sigma_>$ of the damped mode. This enhanced inter-mode coupling provides another energy loss path (in addition to intrinsic damping) to counter the positive rate of work by spin-torques on the destabilized mode, delaying onset of spin-torque-instability and increasing $-I_{eP}^{\text{crit}}$. It is thus not surprising that the $J_{\text{ru}}$-dependence of $-I_{eP}^{\text{crit}}$ closely follows that of $\sigma_>(J_{\text{ru}})$. Further, the broad peaks of $-I_{eP}^{\text{crit}}(J_{\text{ru}})$ in Figs. 4a,b approximately coincide with the quasi-co-resonant condition $\Delta f / \sigma_> < 1$. Finally, since $J_{\text{ru}}^{\max} < J_{\text{ru}}^{\text{device}}$, the broad tail of $\sigma_>(J_{\text{ru}} > J_{\text{ru}}^{\max})$ and the monotonic increase with $t_{\text{FL}2}$ of $J_{\text{ru}}^{\max}$ (e.g., compare Figs. 4a and 4b) yields an explanation for the superlinear increase of $-I_{eP}^{\text{crit}}$ with $t_{\text{FL}2}$.

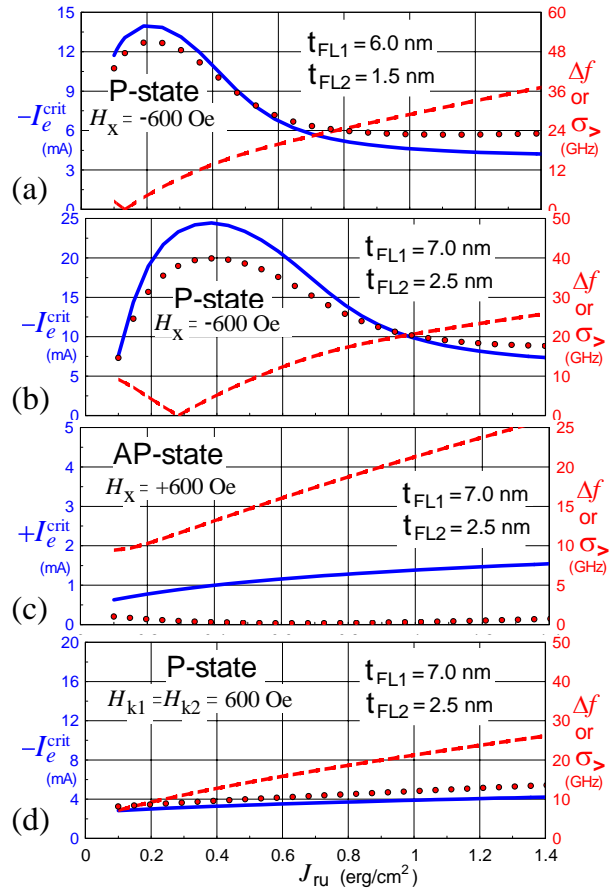

Fig. 4. Modeled values for critical current (solid line), $\Delta f$ (dashed line), and $\sigma_>$ (dotted line) vs $J_{ru}$, for parameter values indicted in text and/or figure. η-coefficients same as used in Fig. 3b.

By contrast, for *positive* $I_{e\text{AP}}^{\text{crit}}$ modeled in Fig. 4c, the co-resonance condition does not occur, and the spin-torque terms actually reduce $\sigma_>(I_e^{\text{crit}})$ *below* that of intrinsic damping ($\approx 4\,\text{Gsec}^{-1}$). Accordingly, $I_{e\text{AP}}^{\text{crit}}$ shows little enhancement with the SFM-FL design, and the model additionally indicates a moderate *reduction* in $I_{e\text{AP}}^{\text{crit}}$ relative to the fictitious case $\eta_{2,3} = 0$ (Fig. 3b). Finally, Fig. 4d shows results for a *bi-stable* SFM-FL with uniaxial anisotropy $H_k = 600\,\text{Oe}$ in FL1,2 replacing the external field (i.e., $H_a \to H_k$ for the $H_5$ term, $H_a \to -H_k$ for $H_7$ in (3)). Although $-I_{eP}^{\text{crit}}(J_{\text{ru}})$ again resembles $\sigma_>(J_{\text{ru}})$ in shape, there is no co-resonance nor superlinear enhancement with $t_{\text{FL}2}$ of $-I_{eP}^{\text{crit}}$, which at $J_{\text{ru}}^{\text{device}} \cong 1\,\text{erg/cm}^2$ is about 10% less than predicted by the $\eta_{2,3} = 0$ model of Fig. 3b. This modeling result is consistent with $H_a = 0$ $dV/dI$ measurements[7] on non-circular devices with magnetostatic shape anisotropy.

The last result in Fig.4d clearly indicates a connection between observable quasi-co-resonant enhancement of $-I_{eP}^{\text{crit}}$, and the presence of an external field *antiparallel* to $\hat{m}_{\text{FL}2}$ (although all states in Fig. 4 are magnetostatically stable at $I_e = 0$ with $J_{\text{ru}} > 0.1\,\text{erg/cm}^2$). This situation would naturally occur in practice for a current-perpendicular-to-plane giant-magnetoresistive magnetic read sensor, where the FL is conventionally stabilized by unidirectional fields from abutted permanent magnet layers.[13] The increase in bias current (while maintaining device stability) afforded by use of the SFM-FL thus has ready application for improving sensor output signal for future read heads in hard disk drives.[7]